\documentclass[prd,amsmath,amssymb,superscriptaddress,preprintnumbers,twocolumn,10pt,nofootinbib,]{revtex4-1}
\pdfoutput=1

\usepackage{graphicx}
\usepackage{dcolumn}
\usepackage{bm}
\usepackage{amssymb}
\usepackage{latexsym}
\usepackage{booktabs}
\usepackage{amsmath}
\usepackage{multirow}
\usepackage{url}

\usepackage{float}
\usepackage[colorlinks=true, linkcolor=red, citecolor=blue]{hyperref}
\usepackage{hyperref}

\usepackage[normalem]{ulem}
\usepackage{color}
\usepackage{array}
\usepackage{enumerate}

\begin{document}

\title{Rapid identification of time-frequency domain gravitational wave signals from binary black holes using deep learning}

\author{Yu-Xin Wang}
\affiliation{Key Laboratory of Cosmology and Astrophysics (Liaoning) \& College of Sciences, Northeastern University, Shenyang 110819, China}
\author{Shang-Jie Jin}
\affiliation{Key Laboratory of Cosmology and Astrophysics (Liaoning) \& College of Sciences, Northeastern University, Shenyang 110819, China}
\author{Tian-Yang Sun}
\affiliation{Key Laboratory of Cosmology and Astrophysics (Liaoning) \& College of Sciences, Northeastern University, Shenyang 110819, China}
\author{Jing-Fei Zhang}
\affiliation{Key Laboratory of Cosmology and Astrophysics (Liaoning) \& College of Sciences, Northeastern University, Shenyang 110819, China}
\author{Xin Zhang}\thanks{Corresponding author}
\email{zhangxin@mail.neu.edu.cn}
\affiliation{Key Laboratory of Cosmology and Astrophysics (Liaoning) \& College of Sciences, Northeastern University, Shenyang 110819, China}
\affiliation{Key Laboratory of Data Analytics and Optimization for Smart Industry (Ministry of Education), Northeastern University, Shenyang 110819, China}
\affiliation{National Frontiers Science Center for Industrial Intelligence and Systems Optimization, Northeastern University, Shenyang 110819, China}

\begin{abstract}

Recent developments in deep learning techniques have provided alternative and complementary approaches to the traditional matched-filtering methods for identifying gravitational wave (GW) signals. The rapid and accurate identification of GW signals is crucial to the advancement of GW physics and multi-messenger astronomy, particularly considering the upcoming fourth and fifth observing runs of LIGO-Virgo-KAGRA. In this study, we used the 2D U-Net algorithm to identify time-frequency domain GW signals from stellar-mass binary black hole (BBH) mergers. We simulated BBH mergers with component masses ranging from 7 to 50 $M_{\odot}$ and accounted for the LIGO detector noise. We found that the GW events in the first and second observation runs could all be clearly and rapidly identified. For the third observing run, approximately 80\% of the GW events could be identified. In contrast to traditional convolutional neural networks, the U-Net algorithm can output time-frequency domain signal images corresponding to probabilities, providing a more intuitive analysis. In conclusion, the U-Net algorithm can rapidly identify the time-frequency domain GW signals from BBH mergers.

\end{abstract}
\maketitle

\section{Introduction}
In 2015, the first detection of a gravitational wave (GW) signal (GW150914) from a binary black hole (BBH) merger initiated a new era of GW astronomy \cite{LIGOScientific:2016aoc}. Furthermore, it provided an important test for the existence of GWs, which were predicted by Albert Einstein in
1916 based on his theory of general relativity \cite{einstein1916naherungsweise}.  On August 17th, 2017, the first binary neutron star (BNS) merger event (GW170817) was detected \cite{LIGOScientific:2017vwq},  which together with its electromagnetic (EM) counterparts began the era of multi-messenger astronomy \cite{LIGOScientific:2017ync}. So far, the LIGO-Virgo-KAGRA collaboration \cite{LIGOScientific:2014pky,VIRGO:2014yos,KAGRA:2018plz} has detected 90 GW
events from compact binary coalescences (CBCs)  \cite{gwcatalog,LIGOScientific:2018mvr,LIGOScientific:2020ibl,KAGRA:2021vkt,LIGOScientific:2021usb}. The study of GW has important implications for fundamental physics, astronomy, and cosmology. For example, GWs can be used to test general relativity \cite{Berti:2015itd,LIGOScientific:2016lio,LIGOScientific:2018dkp,LIGOScientific:2019fpa,LIGOScientific:2020tif,LIGOScientific:2021sio,Gong:2021jgg,Gong:2023ffb}, understand the origins and distributions of astrophysical CBC sources \cite{Mandel:2021smh,vanSon:2021zpk,Broekgaarden:2021efa,Ezquiaga:2022zkx}, and measure cosmological parameters \cite{LIGOScientific:2017adf,Chen:2017rfc,LIGOScientific:2021aug,DES:2019ccw,DES:2020nay,LIGOScientific:2019zcs} (especially providing independent measurements of the Hubble constant using the standard siren method, which is widely discussed in the literature \cite{Holz:2005df,Dalal:2006qt,Nissanke:2009kt,Cutler:2009qv,Camera:2013xfa,Vitale:2018wlg,Bian:2021ini,Cai:2016sby,Cai:2017aea,Cai:2017plb,Zhang:2019ylr,Chen:2020dyt,Gray:2019ksv,Zhao:2010sz,Zhao:2018gwk,Jin:2022qnj,Du:2018tia,Cai:2018rzd,Yang:2019bpr,Yang:2019vni,Bachega:2019fki,Chang:2019xcb,Zhang:2019loq,Mukherjee:2019qmm,He:2019dhl,Zhao:2019gyk,Chen:2020zoq,Mitra:2020vzq,Hogg:2020ktc,Nunes:2020rmr,Borhanian:2020vyr,Jin:2020hmc,Yu:2020vyy,Wang:2021srv,Qi:2021iic,Jin:2021pcv,Zhu:2021bpp,deSouza:2021xtg,Jin:2022tdf,Cao:2021zpf,Leandro:2021qlc,Fu:2021huc,Ye:2021klk,Song:2022siz,Jin:2023zhi,Jin:2023sfc,Jin:2023tou,Han:2023exn,Li:2023gtu}).

Traditional matched-filtering techniques have been greatly successful in the detection of GWs \cite{LIGOScientific:2019ysc} and the performance of matched filtering is considered near-optimal in the confines of LIGO detector noise. Searches are performed with two different latencies: online and offline, as described in Ref.~\cite{Magee:2021xdx}. Online searches are performed nearly in real-time during data collection, whereas offline searches are conducted later using the final, calibrated, and cleaned dataset. Online analysis allows for the rapid release of candidate-related public alerts for multi-messenger counterpart searches. Offline analysis
includes improved background statistics and extensive data calibration, review, and adjustment. Moreover, with relaxed latency requirements, offline analysis can perform more computationally expensive calculations to enhance the separation of signals from the background noise. Owing to these factors, offline analysis is more sensitive than online analysis. Online searches are rapid,
but the ability to separate signals from background noise is not as strong as in offline searches. Furthermore, both types of search demand substantial computational resources. Nevertheless, deep learning can serve as a complementary approach to validate the results generated by the low-latency workflow. In addition, because of the generalization of deep learning \cite{Xia:2020vem}, GW signals beyond the GW template are hoped to be identified, which is expected to be verified by future GW templates. More importantly, deep learning is known for its low computational resource requirements.

Deep learning describes a class of machine learning algorithms. The basic principle of deep learning is to use a multi-layer neural network to gradually extract features from the original input data and make predictions. Thanks to the rapid development of GPU technology, deep learning techniques have gradually been adopted by a wide array of fields in recent years \cite{lecun2015deep,guest2018deep,Baldi:2014kfa,Guest:2018yhq}. The main advantage of using deep learning to identify GW signals is that the algorithm can be pre-trained using a library of known
waveform templates and detector noises. When running an online search, the trained network can be loaded quickly, allowing the rapid and efficient identification of GW sources. 

Recently, GW astronomy based on deep learning algorithms has been intensively discussed in the literature \cite{Wei:2019zlc,George:2017fbn,Chatterjee:2019gqr,Shen:2019ohi,Cuoco:2020ogp,Alvares:2020bjg,Green:2020hst,Green:2020dnx,Dax:2022pxd,Marulanda:2020nww,Singh:2020yau,Mould:2022ccw,Chatterjee:2021lit,McLeod:2022ccr,Langendorff:2022fzq,Chatterjee:2022ggk,Sun:2023vlq,Xiong:2024gpx,George:2016hay,Gabbard:2017lja,Krastev:2020skk,Cabero:2020eik,Jadhav:2020oyt,Xia:2020vem,George:2017pmj,Fan:2018vgw,Wang:2019zaj,Krastev:2019koe,Wei:2020ztw,Verma:2021epx,Moreno:2021fvp,Zhang:2022fwq,Qiu:2022wub,Nousi:2022dwh,Ma:2022esx,Schafer:2022dxv,Trovato:2023bby,Schafer:2021fea}. Among these studies, Refs.~\cite{George:2016hay,Gabbard:2017lja,Krastev:2020skk,Cabero:2020eik,Jadhav:2020oyt,Xia:2020vem,George:2017pmj,Fan:2018vgw,Wang:2019zaj,Krastev:2019koe,Wei:2020ztw,Verma:2021epx,Moreno:2021fvp,Zhang:2022fwq,Qiu:2022wub,Nousi:2022dwh,Ma:2022esx, Schafer:2022dxv,Trovato:2023bby,Schafer:2021fea} focused on the identification of GW signals, with most based on the time-domain analysis. In fact, because the signal strengths of GWs are weak, using 2D data in the analysis of GW detection is sub-optimal \cite{George:2016hay}. However,
many networks work better with 2D images. Hoping to achieve a higher true alarm rate and a lower false alarm rate, analyses are also conducted in the time-frequency domain. Therefore, the identification of the time-frequency domains of GWs using deep learning is also very important, as discussed in Refs.~\cite{Marianer:2020slp,Cuoco:2020ogp,Boudart:2022xib,Ravichandran:2023qma}. The 2D U-Net algorithm \cite{ronneberger2015u} has advantages in image processing, performing well in removing foreground contaminations entangled with the systematic effects of radio
telescopes in neutral hydrogen $21$ cm intensity mapping surveys \cite{Ni:2022kxn,Gao:2022xdb}. 
Given the advantage of U-Net in image processing and the challenges in identifying the extremely faint neutral hydrogen signals and distinguishing GW signals from noise (GW detection and neutral hydrogen $21$ cm intensity mapping surveys both extract weak signals from strong background noise), we are naturally motivated to explore the potential of the 2D U-Net algorithm in identifying time-frequency domain GW signals from BBHs.

In this article, we explore the potential of the U-Net algorithm in identifying GW signals. Unlike traditional convolutional neural network (CNN) algorithms, U-Net can output images corresponding to probabilities, enabling a more intuitive investigation of the time-frequency-domain signal. Our goal was to train the network to quickly identify GW signals in the time-frequency domain. Given that the currently detected GW signals are mainly BBH mergers, we focused on the identification of BBH mergers. We began by training the network using simulated BBH signals and subsequently applied the trained network to real observations. Moreover, this is the
first study to perform time-frequency-domain analysis and apply the network to the O3 observations. Through this study, we aim to shed light on the possibility of using U-Net to identify the GW signals from BBH mergers. Neutron star–black hole (NSBH) and BNS mergers have been omitted from this study but will be included in future research.

This article is organized as follows: In Sec.~\ref{sec2}, we introduce the methodology used in this study. In Sec.~\ref{sec3}, we report the identification results of the trained network
with the simulation and O1, O2, and O3 data. The conclusion is given in Sec.~\ref{sec4}.

\section{Methodology}\label{sec2}
\subsection{Dataset assembly}\label{sec2.1}
In this study, we focused on the GW signals produced by BBH mergers. We generated six sets of data for training six models, each containing 170000 samples of pure background noise and 170000 samples of mixed GW signal and background noise, for different observation runs and detectors. For the simulated GW signals, we adopted the waveform model $\rm{IMRPhenomXPHM}$ \cite{Pratten:2020ceb} generated by $\tt PyCBC$ \cite{alex_nitz_2023_7885796}. For the background noise, we selected segments from the publicly available O1, O2, and O3 data of LIGO Hanford and LIGO Livingston, wherein no GW events were detected.

The synthetic data was described by
\begin{equation}
s(t)=h(t)+n(t),
\end{equation}
where $h(t)$ is the GW signal and $n(t)$ is the background noise.

The duration of the training dataset was 8 s (we set the merger-time range for the GW signal to $[6.5, 7.5]$ s) and the sampling rate was 4096 Hz. All the data were whitened and passed through a band-pass filter with a frequency in the range of $[30, 900]$ Hz. Owing to the edge
effect of the band-pass filter, the data in the edge 0.25 s were set to zero. Subsequently, we utilized the short-time Fourier transform and applied a Hanning window to transform the signals into the time-frequency domain for analysis. Specifically, the window length was set to 0.1 s and the overlap was half the window length. We preprocessed our data to fit the required network dimensions, setting both the time and frequency bins to 160. Subsequently, we applied maximum normalization to each image by dividing all pixel values by the highest image pixel value, ensuring the maximum pixel value in each image was normalized to 1. The simulated GW parameters are shown in Table~\ref{tab1}. The spin magnitudes were uniformly distributed in the range of $[0, 0.99]$.

\begin{table}[!htp]
\caption{Distribution of simulated GW waveform parameters. Parameters not listed were set to zero for simplicity.}\label{tab1}
\centering
\setlength\tabcolsep{18pt}
\renewcommand{\arraystretch}{1.5}
\begin{tabular}{cc}
\hline \hline Parameter & Uniform distribution \\
\hline Component masses & $m_1,m_2 \in[7,50]~ M_{\odot}$ \\
Coalescence phase & $\Phi \in[0,2 \pi]~ $rad \\
Polarization  & $\Psi \in[0,2 \pi]~ $rad \\
Inclination & $\cos\iota \in[-1,1]$ \\
Declination & $\sin\theta \in[-1,1]$ \\
Right ascension & $\varphi \in[-\pi, \pi]~ $rad \\
Chirp distance & $\mathrm{d_{c}^{2}}\in[130^{2},350^{2}]~ \mathrm{Mpc^{2}}$\\
\hline \hline
\end{tabular}
\end{table}

Here we emphasize that the detections of BNS mergers are also important, particularly for observations of the follow-up electromagnetic counterparts, which could greatly advance the fields of multi-messenger astronomy and standard siren cosmology. In this study, we only considered BBH mergers for two reasons. First, to ensure the robustness of the algorithm, as many verification sets as possible were needed. Given that the CBCs currently detected by LIGO-Virgo-KAGRA are mainly BBH mergers, they serve as suitable candidates for testing and validating our methods. Second, owing to the extended duration of BNS merger detections compared to BBH mergers, the data dimension is relatively high. Without employing PCA for dimensionality reduction, the dataset requirements would experience exponential growth; consequently, BNS mergers should be treated with dedicated networks and separately from BBH mergers \cite{Qiu:2022wub,Baltus:2022pep}.

\subsection{U-Net architecture}

\begin{figure*}[!htbp]
\includegraphics[width=0.8\textwidth]{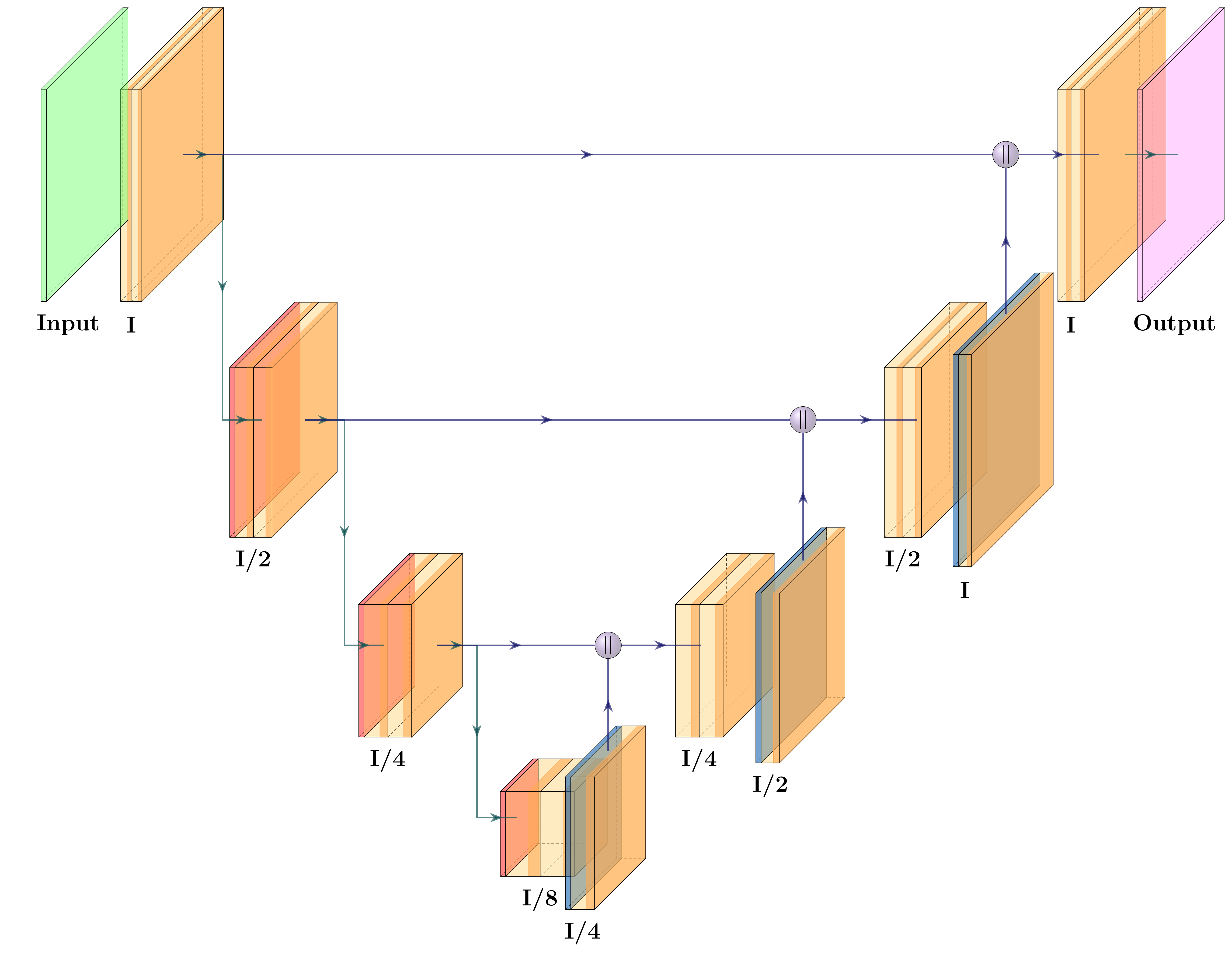}
\centering
\caption{\label{fig1} Training of a CNN with U-Net architecture. Here each color represents a structure in the U-Net network: yellow cubes represent the convolutional layers and ReLU sections, red cubes represent pooling layers in the down-sampling, blue cubes represent the transposed convolutional layers, and grey cubes represent connection layers. This visualization was made using the PlotNeuralNet library \cite{PlotNeuralNet-web}.}
\end{figure*}

\begin{figure*}[!htp]
\centering
\includegraphics[width=0.8\textwidth]{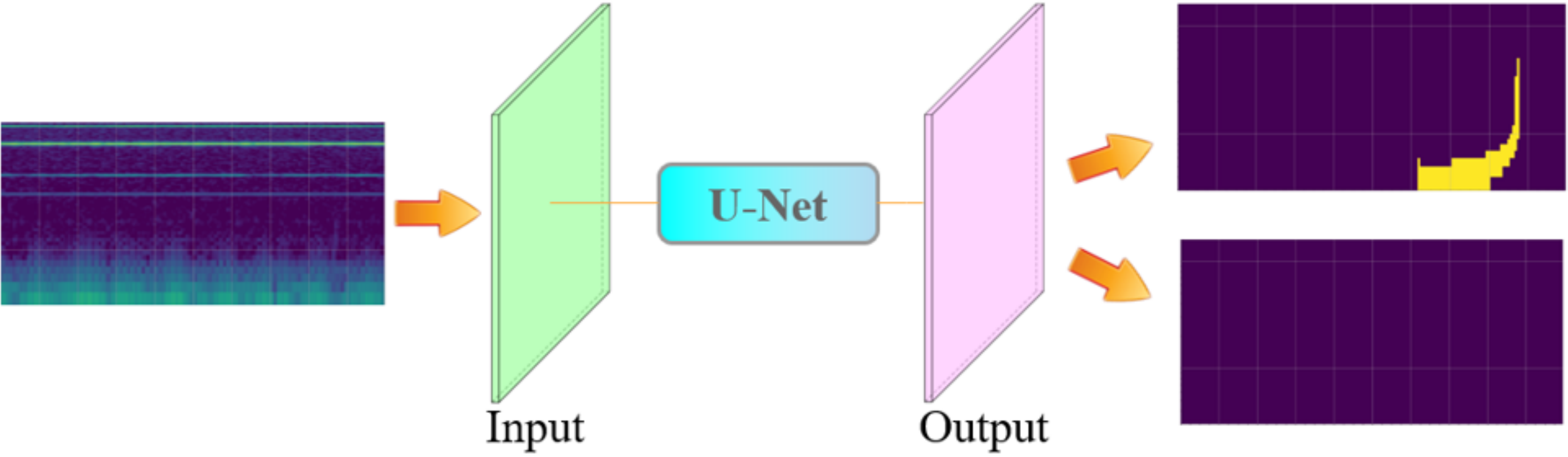}
\centering \caption{\label{fig2} Flowchart of the network used in this study. The observed signal \cite{gwpy}  (mixed GW signal and noise or noise only) is processed by the U-Net algorithm. If a GW signal is present, a GW signal image will be output. If there is no signal, an empty signal image will be output.}
\end{figure*}

The typical use of convolutional networks is on classification tasks, where the output of the image is a single class label. However, in many visual tasks, assigning a single class label to an entire image is not sufficient. Instead, the desired output should include pixel-level localization, where a class label is assigned to each individual pixel. The U-Net algorithm can achieve this through semantic segmentation, which is an approach for identifying the class of an object for each pixel. This approach is particularly useful for tasks such as object detection,
where the precise location and shape of the object of interest need to be identified. 

\begin{table*}[!htp]
\caption{Description of the hyperparameters in the U-Net architecture design.}\label{tab2}
\centering
\renewcommand{\arraystretch}{1.5}
\begin{tabular}{cccc}
\hline\hline Hyperparameter & Description & Prior value & Optimum value \\
\hline$\eta$ & learning rate for optimizer & $10^{-3}, 10^{-4}, 10^{-5},10^{-6}$ & $10^{-5}$ \\
\hline$\omega$ & weight decay for optimizer & $10^{-4}, 10^{-5}, 10^{-6}, 10^{-7}$ & $10^{-5}$ \\
\hline$n_{\text {filter }}$ & initial number of convolution filters & $16,32$ & 32 \\
\hline$b$ & batch size & 32,64 & 32 \\
\hline$\Omega$ & optimizer for training & Adam, NAdam & NAdam \\
\hline\hline
\end{tabular}
\end{table*}

The U-Net network is a CNN originally developed for biomedical image segmentation \cite{ronneberger2015u}. Although based on a CNN, it has undergone significant structural modification. Unlike standard CNN architecture, U-Net includes many feature channels in the upsampling,\protect\footnote{It refers to any technique that allows an image to be changed to a higher resolution. The simplest way is resampling and interpolating: rescale the input image to a desired size, count the pixels at each point, and use interpolation methods such as bilinear interpolation to interpolate the remaining points to complete the upsampling process.} allowing the network to propagate contextual information to higher resolution layers through a series of transpose convolutions.\footnote{Each convolutional layer is composed of a specified number of kernels. Each kernel multiplies the input feature values with weights and adds the biases to obtain outputs. Different kernels get different parameter values after training.} The main idea behind U-Net is to add successive layers to the traditional contracting network, replacing the
pooling operation with an upsampling operation. This leads to an increase in output resolution, as these layers produce a U-shaped structure that is almost symmetric with the contracted part.

The U-Net architecture is illustrated in Fig.~\ref{fig1}. It consists of a contracted path (left side) and an extended path (right side). The contracting path follows the typical architecture of a convolutional network. It comprises the repeated application of $3\times3$ convolutions (unpadded convolutions), each followed by a rectified linear unit (ReLU) and a $2\times2$ max pooling\footnote{{This layer scans the data according to a specified stride within a window of a certain length. Then, it outputs the maximum value of the data in each scanning window.}} operation with stride 2 for downsampling. At each downsampling step, the number of feature channels is doubled. Every step in the expanding path consists of an upsampling of the feature map followed by a $2\times2$ convolution that halves the number of feature channels, a concatenation with the corresponding cropped feature map from the contracting path, and two $3\times3$ convolutions, each followed by a ReLU. The cropping is necessary due to the loss of border pixels in each convolution. In the final layer, a $1\times1$ convolution is used to map each 64-component feature vector to the desired number of classes. In total, the network has 3
convolutional layers.

The network flowchart is shown in Fig.~\ref{fig2}. Whenever a time-frequency-domain image is sent to the network, it outputs a noise-reduced image. If the image includes a GW signal, the signal will be highlighted; otherwise, the image will appear almost blank, with pixel values close to
zero, rather than exactly zero. For large test datasets, analysis is performed by extracting the maximum pixel value from each image and comparing it with a predetermined threshold. Images with maximum values exceeding this threshold are considered to contain a GW signal, whereas those below are deemed signal-free. We included sigmoid activation in the output layer of the network so that the output of the network is in the range [0, 1], and a threshold of 0.5 was set.

\subsection{Training}

During the training process, the coefficients of neural network are determined. To assign initial random values to CNN parameters, we used the ``Xavier'' initialization, which is designed to keep the scale of gradients roughly the same in all layers. Then, we used the binary cross-entropy loss function to evaluate the deviation between the predicted and the actual values in the training data.

The key component of a CNN is the convolutional layer, which applies a set of filters to the input. The network consists of a series of stacked layers. In the first convolutional layer, we set the number of convolution kernels to 32. The kernel size determines the convolutional field of view and is fixed at $3 \times 3$. To maintain the output dimensionality, we employed the same padding method for both convolutions and transpose convolutions to handle sample boundaries. The stride determines the kernel traversal step size on the images. We used the default stride settings of 1 for convolutions and 2 for transpose convolutions.

Our selected U-Net architecture was trained end-to-end for signal identification using the simulated data introduced in Sec.~\ref{sec2.1}. The details of the hyperparameters used in this work are listed in Table~\ref{tab2}. The prior values are those we used during the training process that can achieve a high true alarm rate with a threshold of 0.5, and the optimum values are those we finally selected, with the best ROCs. The NAdam optimizer\footnote{{In the process of deep learning backpropagation, the optimizer guides each parameter of the loss function (objective function) to update the appropriate size in the right direction, so that the updated parameters make the value of the loss function (objective function) approach the global minimum continuously.}} was used in the analysis with the default TensorFlow parameters \cite{reddi2019convergence}. The hyperparameters were carefully fine-tuned to optimize the network. The batch size and number of initial convolution filters were optimized to 32, both of which were limited by the GPU memory. The learning rate\footnote{{In machine learning and statistics, the learning rate is a tuning parameter in an optimization algorithm that determines the step size at each iteration while moving toward a minimum loss function.}} was set to $10^{-5}$ and the weight decay to $10^{-4}$. We set the dropout rate to 0.2 and used batch normalization\footnote{{In the training process of each batch after the data passes through the activation layer, the activation value of each batch of data is normalized. In other words, the average value of the sample data of each batch is normalized to 0 and the variance is normalized to 1. The purpose of this step is to make the result of batch normalization the same as the original input data, which maintains the possibility of retaining the original structure.}} to maintain the mean and variance of the input data distribution of each layer in the U-Net within a certain range. The total number of trainable parameters was $3.1\times10^6$. We applied a ReLU activation in every convolution. Finally, we adopted the 200 epoch calculation scheme to improve our results.

At the end of every epoch, the network training performance was evaluated using the average accuracy of the validation set. The training process was completed within 6 h on four NVIDIA GeForce RTX A6000 GPUs, each with 48 GB of memory.

The identification time for 173 GW signals\footnote{{Note that we consider the GW signals from both LIGO Hanford and LIGO Livingston. We consider 86 GW signals for LIGO Hanford and 87 signals for LIGO Livingston.}} was 0.1868 s, meaning that the identification time for each GW signal was approximately 1 ms. Thus, compared to matched filtering, U-Net can save computation resources.

Consistent with the method used in Ref.~\cite{Schafer:2022dxv}, the effective maximum luminosity distance was selected based on the chirp mass; the smaller the chirp mass, the smaller
the maximum luminosity distance at which injections were placed. The chirp distance is given by \cite{LIGOScientific:2007npa}
\begin{equation}
d_c = d \left( \frac{\mathcal{M}_{c,0}}{\mathcal{M}_c} \right)^{5/6},
\end{equation}
where $d$ is the luminosity distance, $\mathcal{M}_c = \left( m_1 m_2 \right)^{3/5} / (m_1 + m_2)^{1/5}$ is the chirp mass, and $\mathcal{M}_{c,0} = 1.4/2^{1/5} \, M_\odot$ is a fiducial chirp mass used as a basis for calculation.

When considering a fixed chirp mass, the injections are uniformly distributed within the sphere of the adjusted maximum luminosity distance. The formula we used to estimate the sensitive volume can be written as
\begin{equation}
V(\mathcal{F}) \approx \frac{V(d_{\text{max}})}{N_I} \sum_{i=1}^{N_{I,\mathcal{F}}} \left( \frac{M_{c,i}}{M_{c,\text{max}}} \right)^{5/2},
\end{equation}
where $d_{\text{max}}$ is the maximum luminosity distance, $V(d_{\text{max}})$ is the volume of a sphere with radius $d_{\text{max}}$, $N_{I,\mathcal{F}}$ is the number of found injections at a FAR of $\mathcal{F}$, $N_I$ is the total number of injections performed, $\mathcal{M}_{c,i}$ is the chirp mass of the $i$th found injection, and $\mathcal{M}_{c,\text{max}}$ is the upper limit on the injected chirp masses.

Before proceeding to the results and discussion, we should note that we trained 6 models based on different noise sources. When analyzing a specific observation run of a specific detector, we used the corresponding model for prediction. We combined the network outputs by adding the outputs of LIGO Hanford and LIGO Livingston at the same moment. Because the output range of a single network is [0, 1], the combined output range is [0, 2]. The advantage of combining the outputs of two detectors in this way is that, when facing glitches, even if the output of one detector indicates a signal, that of the other is unlikely to indicate a signal at the same time.
This increases the robustness of the system to glitches, because they typically do not occur at the same time in both detectors and have similar characteristics.

\section{Results and discussion}\label{sec3}
\subsection{Simulation results}

\begin{figure}[!htp]
\includegraphics[width=0.5\textwidth]{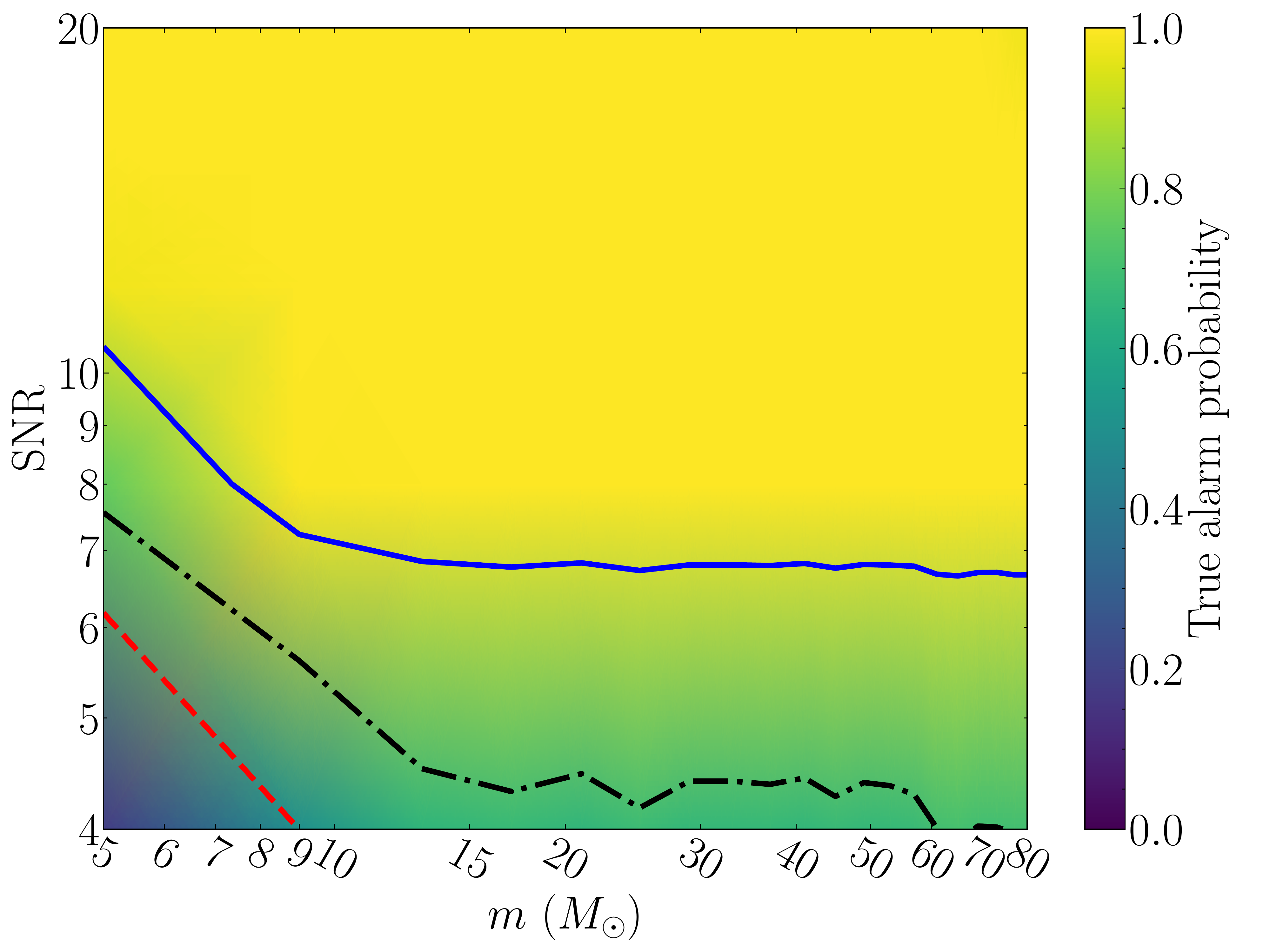}
\centering \caption{\label{fig3} Network robustness on $m$ with respect to the SNR. Note that only equal-mass BBH mergers were considered. The dashed red line is the 50\%  contour line, the
dot-dashed black line is the 70\% contour line, and the solid blue line is the 90\% contour line.}
\end{figure}

\begin{table}[H]
\caption{Testing the robustness of networks to noise with GW signals at different times.}\label{tab3}
\centering
\setlength\tabcolsep{36pt}
\renewcommand{\arraystretch}{1.5}
\resizebox{\linewidth}{!}{%
\begin{tabular}{cc}
\hline \hline   Time (s) & True alarm probability \\
\hline 6.75 & 0.9722 \\
6.90 & 0.9820 \\
7.00 & 0.9942 \\
7.10 & 0.9896 \\
7.25 & 0.9834 \\
\hline \hline
\end{tabular}%
}
\end{table}

For the first and second tests, we used the network trained on the O3 data from LIGO-Hanford. We first examined the GW-detection ability of the network for a single mass $m$ and SNR. In the training dataset, we considered equal-mass BBH mergers; the BH mass was sampled in the range of $[5, 80]~M_{\odot}$, and SNR in the range of $[4, 20]$. After 100 epochs of training, the accuracy of the test dataset was 90\%. To illustrate the capability of the network, Fig.~\ref{fig3} shows the results with $m$ and SNR in the ranges of $[5, 80]~M_{\odot}$ and $[4, 20]$, respectively. The network performed well in the high-mass and high-SNR regions, but not in the low-mass and low-SNR regions. Therefore, for the lower-SNR region ($4<\text{SNR}<5$), the the network also performed poorly. The primary cause for this is that the network is more effective at handling short signals. As the chirp mass decreases, the GW signals last longer and become buried within the characteristics of the random noise. In other words, because the SNR is integrated throughout the GW signals, at the same SNR, short signals have higher strain values than those of long signals. Therefore, it is easier to identify the short signals. For the high-mass and low-SNR regions, the network performs better. The network can achieve an accuracy of 90\% when the single mass is larger than $9~M_{\odot}$ and SNR is greater than 8.

\begin{figure}[htbp]
\includegraphics[width=0.5\textwidth]{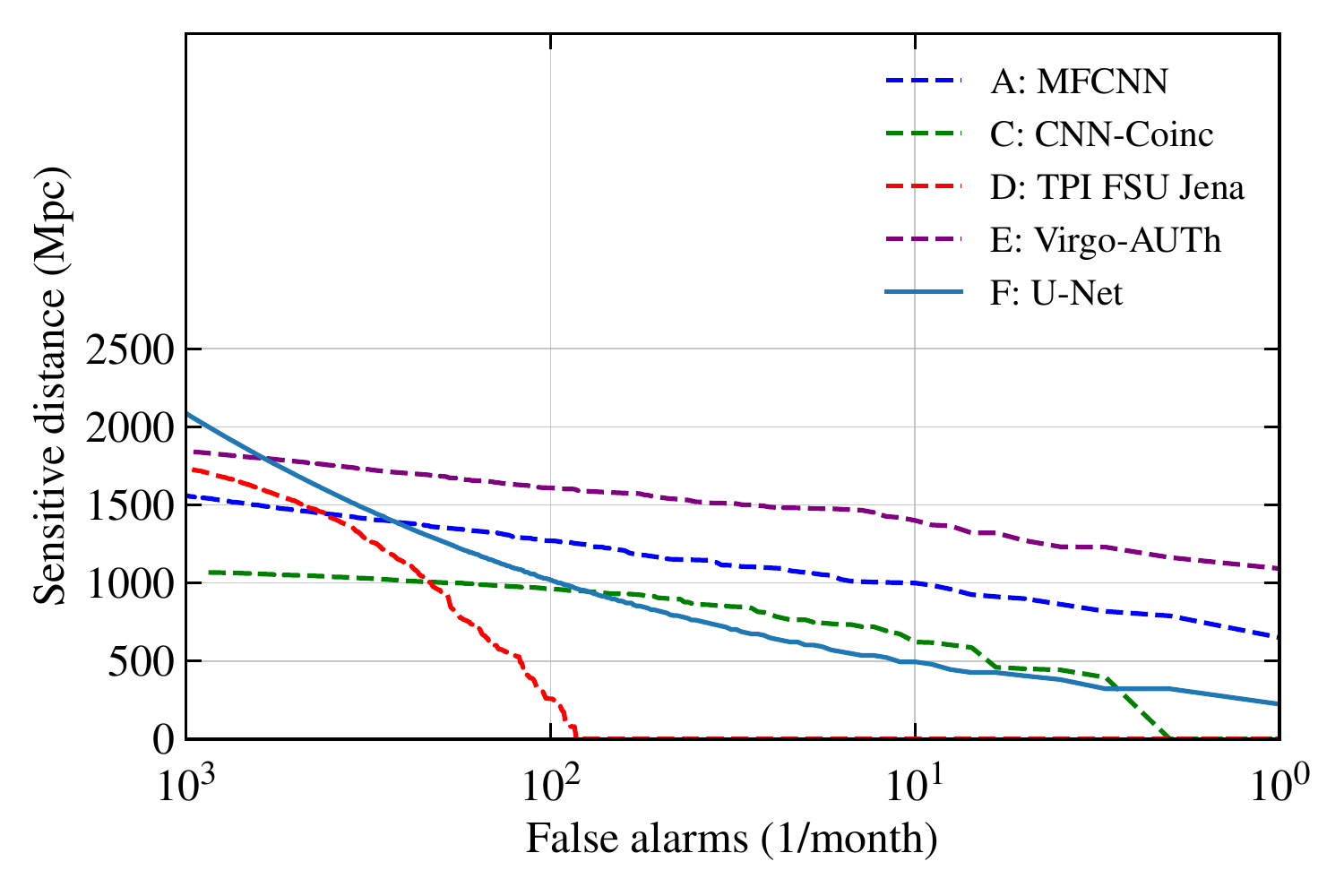}
\centering \caption{\label{fig4} Sensitive distances of U-Net and other machine learning methods \cite{Schafer:2022dxv} as functions of the FAR. UNet is shown with a solid line, other methods with dashed lines. The FAR was calculated on a background set that does not contain any injections.}
\end{figure}

Second, we tested the robustness of networks where the GW signals were added to noise at different times. We selected several fixed signal added times, $[6.75, 6.9, 7, 7.1, 7.25]$ s, to test whether the network can identify signals with different added times. We generated nearly 20,000 samples. The true alarm probabilities are listed in Table~\ref{tab3}. We used the maximum pixel value in the image predicted by the network to compare with a threshold set to 0.5. We also generated nearly 20,000 pure-noise samples with a false alarm probability of 0.0026.

For the final test, we used two models trained separately on the O3 data from LIGO Hanford and LIGO Livingston. We used O3a noise for evaluation and downloaded six days of data starting from June 6, 2019 (GPS start time 1243877376 and GPS end time 1244340224). All noise was given the “data" quality flag, and none of the “CBC\_CAT1,” “CBC\_CAT2,” or “CBC\_HW\_INJ” flags were active. We considered only segments where the data from both LIGO Hanford and LIGO Livingston
met these conditions, with a usable data duration of 3.25 d. We shifted the data from LIGO Livingston by 100 s ten times, while keeping the data from LIGO Hanford fixed. This allowed us to generate 32.5 d of coincident real noise for both detectors.

We sampled the injections from the distributions defined in Table~\ref{tab1} using the chirp distance. The spin magnitudes were uniformly distributed in the range of $[0, 0.99]$. The merger time of two subsequent signals was set at 20 s to avoid any overlap.

In Fig.~\ref{fig4}, we show the variation in the sensitive distances of different machine learning methods with the number of sources that can be detected at a particular FAR. Consistent with Ref.~\cite{Schafer:2022dxv}, the FAR is plotted at 1000 per month. Compared to the Virgo-AUTh \cite{Nousi:2022dwh} and the MFCNN \cite{Wang:2019zaj}, U-Net can reach a higher sensitive
distance at a FAR of 1000 per month. However, as the FAR decreases, the sensitive distance of U Net drops rapidly. Compared to the CNN-Coinc \cite{Gabbard:2017lja,Schafer:2021cml}, U-Net performs better at FARs of 1000, 100, and 1 per month. Within the mass range of currently detected BBH events, machine learning methods often achieve comparable or superior sensitivity distances to matched filtering at FARs significantly higher than 1 per year \cite{Marx:2024wjt}.  In conclusion, U-Net is not significantly superior to other networks but is strongly competitive.

\subsection{Robustness test of U-Net}
\begin{figure*}[!htbp]
\centering
\includegraphics[width=1\textwidth]{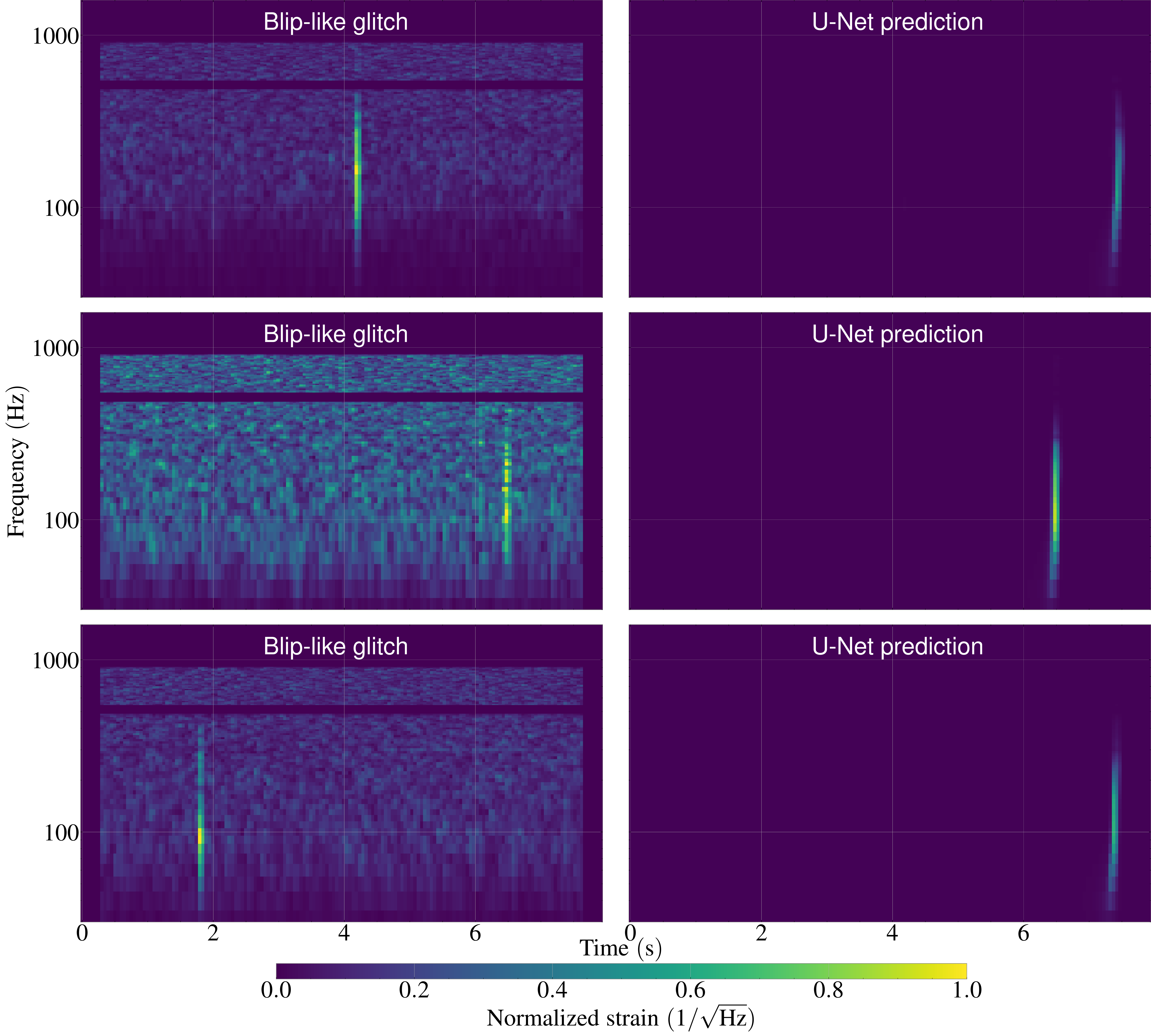}
\centering \caption{\label{fig5} {Blip-like glitches at different times (left) and the prediction results of U-Net (right).}}
\end{figure*}

In the process of identifying GW signals, as typical non-stationary noise in GW detectors, glitches often have a large impact on the matched-filtering method. Glitches may also have an impact on the false alarm rate of U-Net. Hence, in this subsection, we report a test on the impact
of glitches on the false alarm rate of a network that has not been trained on glitches.

We chose blip glitches to represent the glitches. Blip glitches are very short-duration transients ($\mathcal{O}(10)$ ms) and have a wide frequency bandwidth ($\mathcal{O}(100)$ Hz) \cite{Cabero:2019orq}. We chose blip glitches for two reasons, (i) they are common in the GW data, and (ii) in the time-frequency domain they appear similar to high-mass BBH mergers, and thus can be mistaken for a real GW signal \cite{Cabero:2019orq,Nitz:2017lco}.

We used the network trained on the O3 data from LIGO-Hanford to identify the blip glitches. We selected blip-like glitches from the LIGO data, then placed the glitches at different times and used the trained network to identify them. As mentioned above, if the network does
not identify the signal, an empty signal image will be output. We randomly tested 200 glitches, with ten identified as signals. Thus, the false alarm rate was approximately 5\% for the blip-like glitch. In Fig.~\ref{fig5}, we show three typical blip-like glitches at different times and the predictions of U-Net. The blip glitches are similar to the waveform of a GW at the merge-time, and both resemble the shape of a chirp signal. This might be the reason behind the high false alarm rate.

\subsection{Application to the real observations}

\begin{figure*}[!htbp]
\centering
\includegraphics[width=1\textwidth]{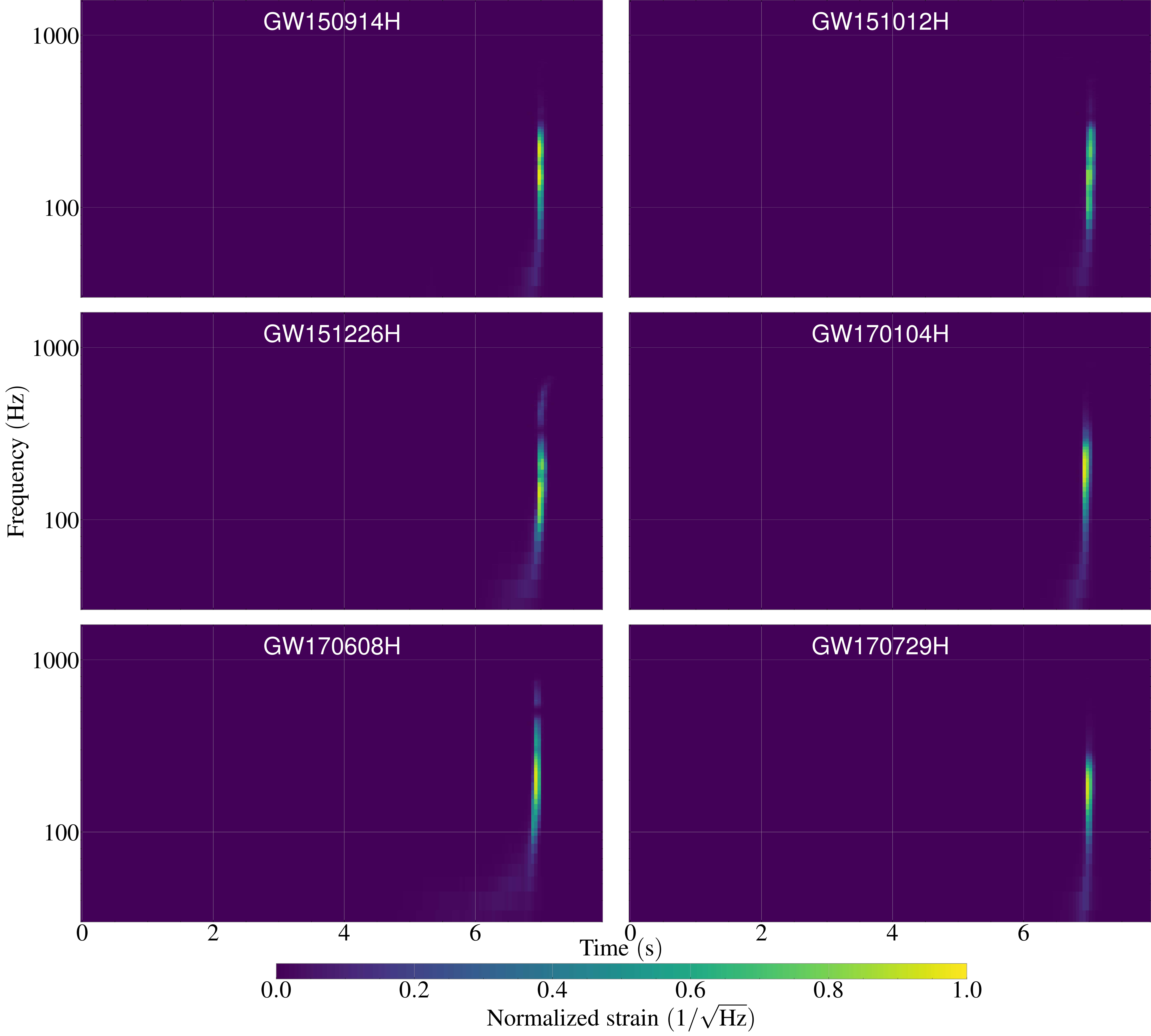}
\centering \caption{\label{fig6} Time-frequency representations of the GW signals of O1 and a portion of O2, identified by the trained network (the remaining O2 signals are presented in Fig.~\ref{fig7}). In the event names, H and L represent the signals from LIGO Hanford and LIGO Livingston, respectively.}
\end{figure*}

\begin{table}[H]
\caption{GW events not considered in this study.}\label{tab4}
\centering
\setlength\tabcolsep{36pt}
\renewcommand{\arraystretch}{1.5}
\resizebox{\linewidth}{!}{%
\begin{tabular}{cc}
\hline \hline Name & Event type \\
\hline GW170817 & BNS \\
GW190425 & BNS \\
GW190814 & NSBH \\
GW190917\_114630 & NSBH \\
GW191219\_163120 & NSBH \\
GW200115\_042309 & NSBH \\
GW200210\_092254 & NSBH \\
\hline \hline
\end{tabular}%
}
\end{table}

\begin{figure*}[!htbp]
\includegraphics[width=\textwidth]{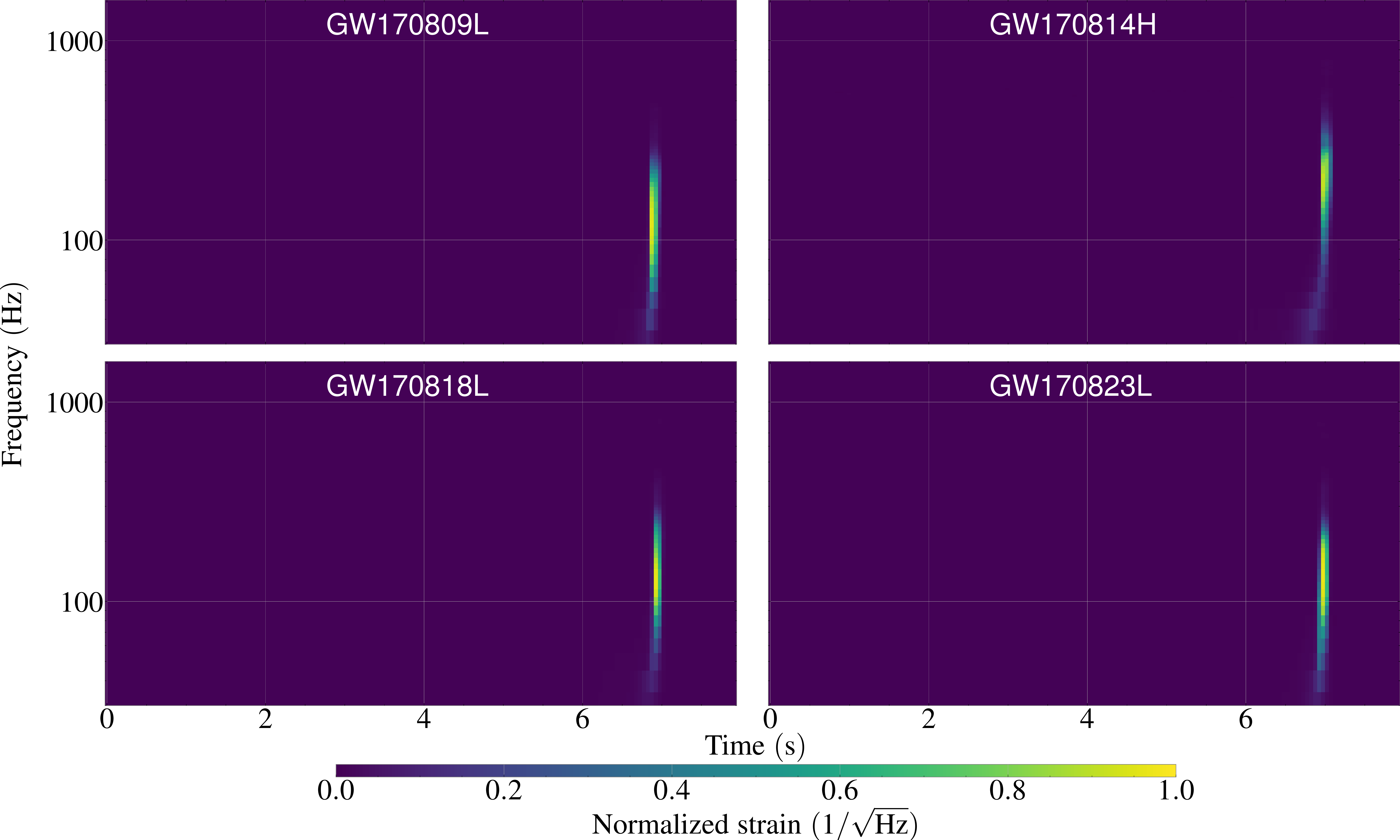}
\centering \caption{\label{fig7} Same as Fig.~\ref{fig6}, but for the remaining time-frequency representations of the GW events from GW170809 to GW170823.}
\end{figure*}

\begin{figure*}[!htbp]
\includegraphics[width=\textwidth]{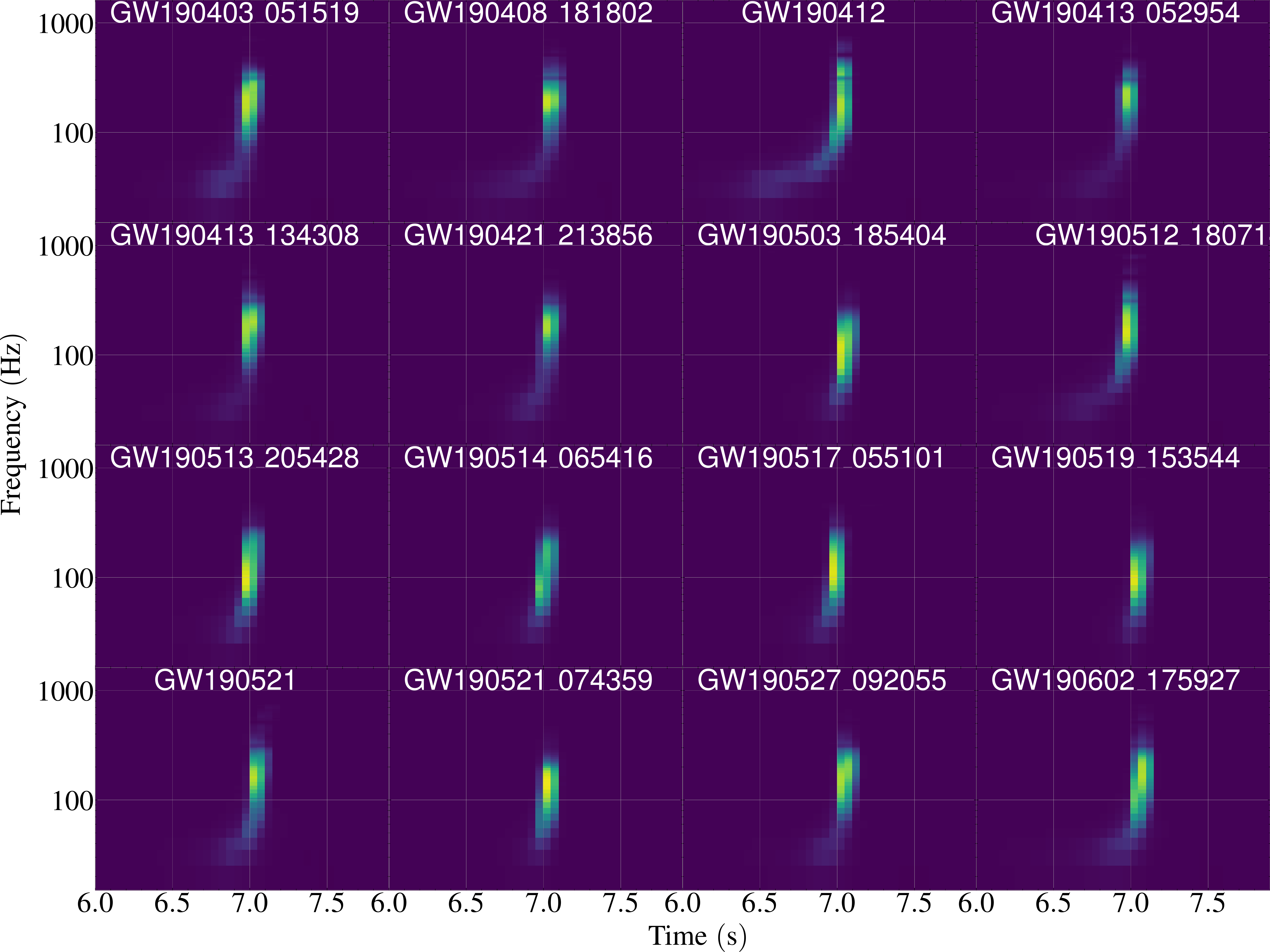}
\centering \caption{\label{fig8}  Time-frequency representations of a portion of the GW signals of O3 identified by the trained network (the remaining O3 signals are presented in Figs.~\ref{fig9}--\ref{fig11}). The colorbar settings are consistent with Fig.~\ref{fig6}.  For clarity, only the GW signals from 6 s to 8 s are shown. For each GW signal, only the better identified signal from LIGO Livingston or LIGO Hanford is shown for simplicity.}
\end{figure*}

\begin{figure*}[!htbp]
\includegraphics[width=\textwidth]{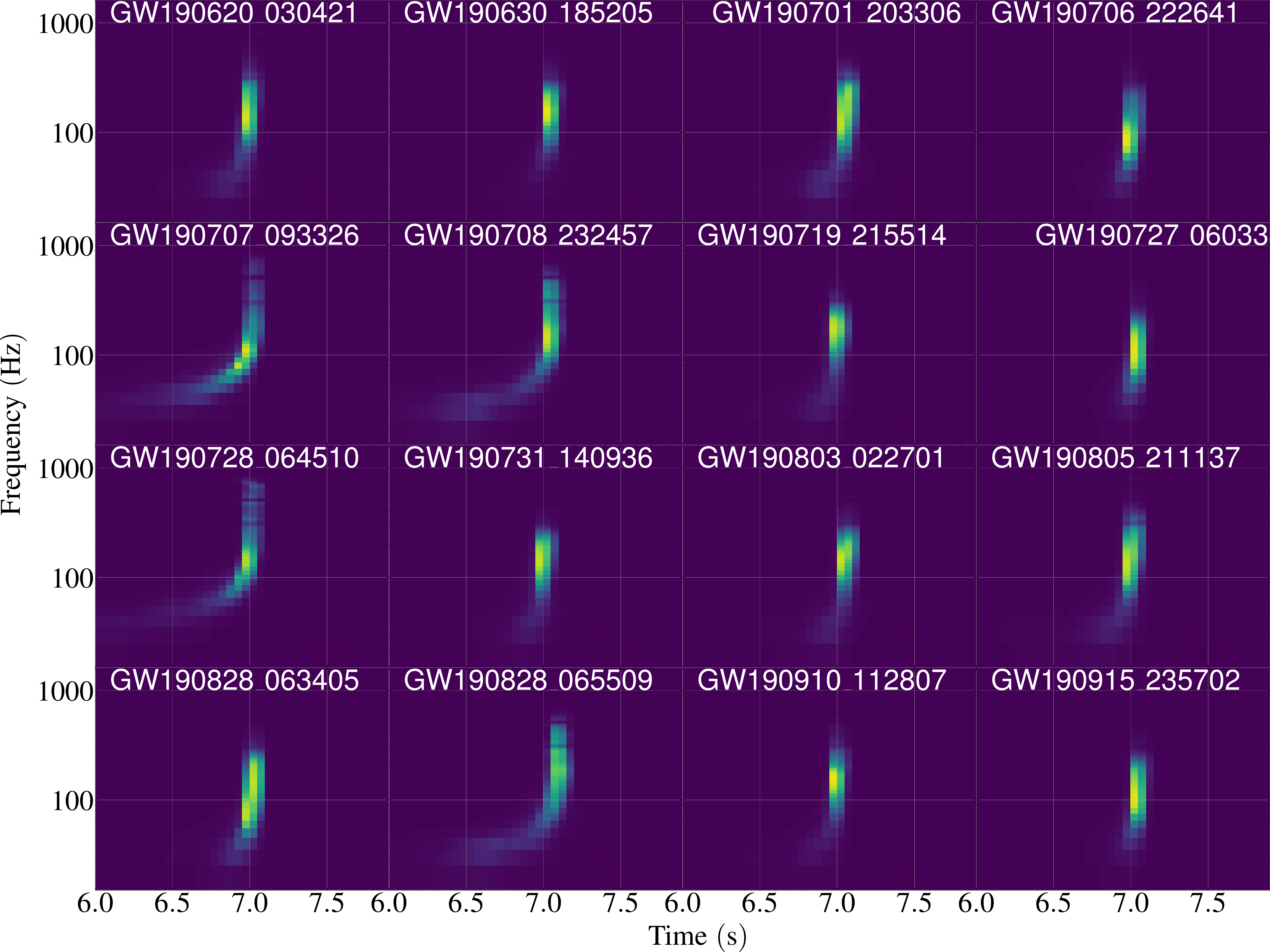}
\centering \caption{\label{fig9} Same as Fig.~\ref{fig8}, but for the remaining time-frequency representation of GW events, from GW190620$\_$030421 to GW190915$\_$235702.}
\end{figure*}

\begin{figure*}[!htbp]
\includegraphics[width=\textwidth]{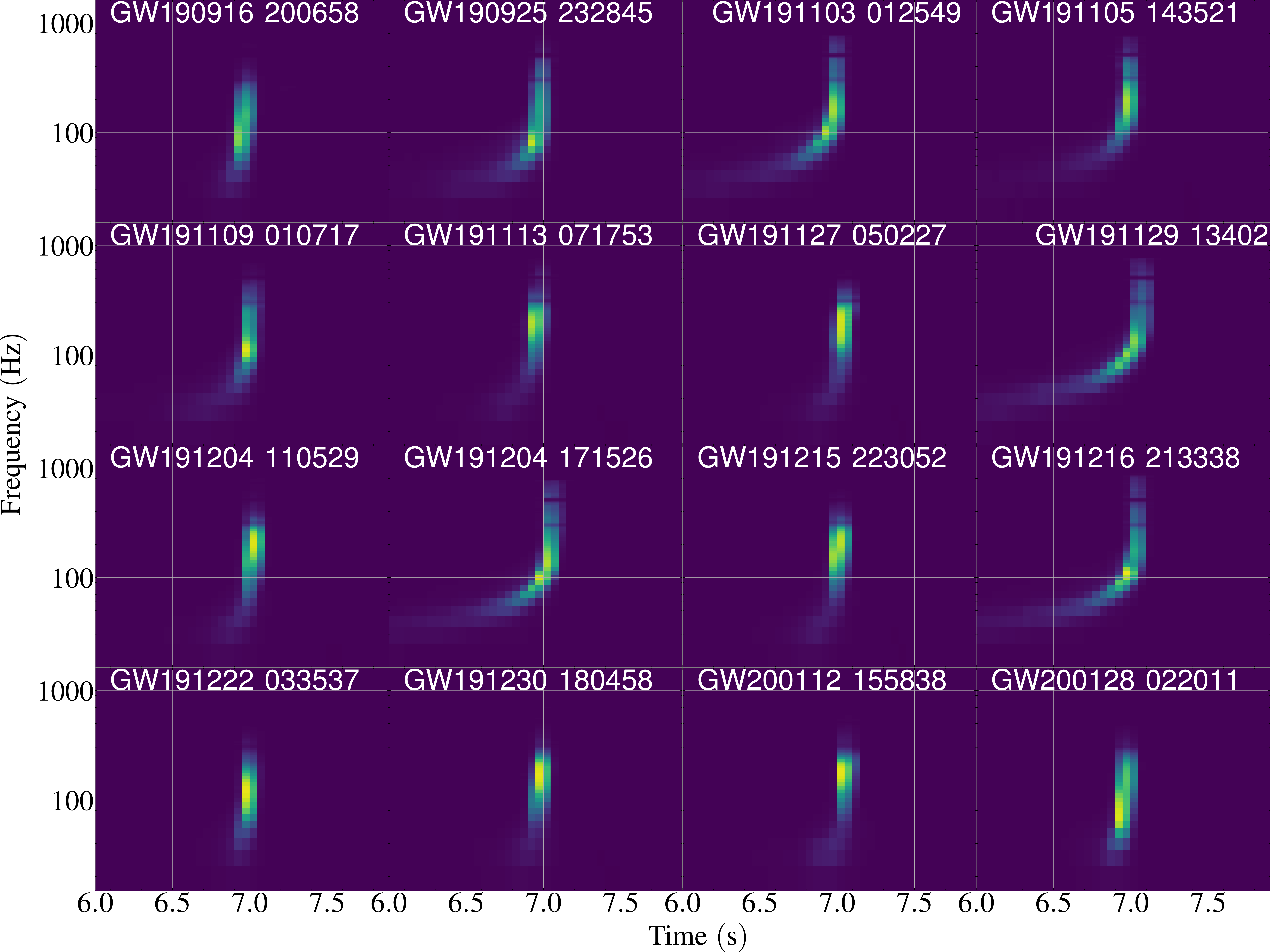}
\centering \caption{\label{fig10}  Same as Fig.~\ref{fig8}, but for the remaining time-frequency representation of GW events, from GW190916$\_$200658 to GW200128$\_$022011}.
\end{figure*}

\begin{figure*}[!htbp]
\includegraphics[width=\textwidth]{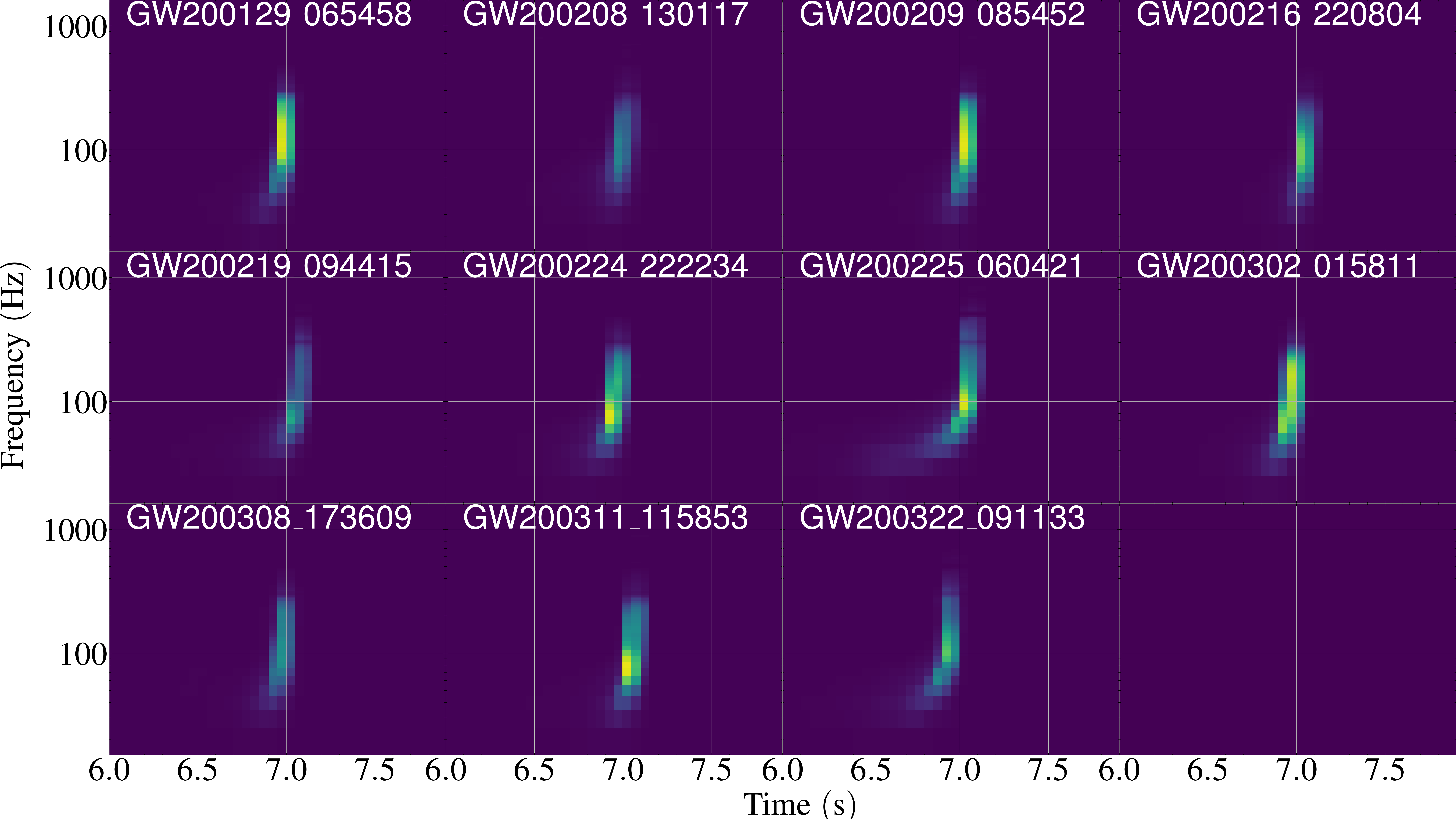}
\centering \caption{\label{fig11}  Same as Fig.~\ref{fig8}, but for the remaining time-frequency representation of GW events, from GW200129$\_$065458} to GW200322$\_$091133.
\end{figure*}

For real observations, when analyzing a specific observation run from a specific detector, we used the corresponding model for prediction. In this subsection, we report the identification results. We tested the real GW events by placing the signal at approximately 7 s to make
it easier to confirm the results. For each GW event, we identified the signals from LIGO Hanford and LIGO Livingston. Owing to the low SNR of some GW signals, there will be missed detections during identification. Hence, we utilized the union of LIGO Hanford and LIGO Livingston, i.e., we considered the signal identified as long as it was detected by either of the two detectors. Note that in the present study, we only considered the identification of GW signals from BBH mergers; GW events involving NSs were not considered (summarized in Table~\ref{tab4}).

In Fig.~\ref{fig6} and \ref{fig7}, we show the time-frequency representation of the GW signals of O1 and O2 identified by the trained network. All the GW events were identified by the network. Further, we tested the network by attempting to identify BNS mergers and found that the network is unable to identify them, which is consistent with our expectations. From the figure, we can also see that the identified signals obtained from the network are different for different GW events. This is primarily because the time-frequency signals of GWs for different chirp masses are different. This also means that the time-frequency representation of GW signals identified by U-Net may be used to preliminarily determine the chirp masses of the GW sources. Previous research shows that prior selection of the GW parameter has an important impact on the Bayesian inference \cite{Vitale:2017cfs}. Our results show that the network can not only rapidly and accurately identify the GW signals but also aid with the later Bayesian inferences.

Subsequently, we applied the trained network to identification in the O3 data. In Figs.~\ref{fig8}--\ref{fig11}, we show the timefrequency representations of the GW signals of O3 identified by the trained network. The trained network shows a strong ability to identify GW signals. Fifty-nine GW signals were clearly identified by the network. Note that for O3 we considered a total of 73 GW signals for analysis and only show the identified results. Approximately 80\% (80.8\%) of GW signals in O3 could be rapidly and accurately identified.

\section{Conclusion}\label{sec4}

In this study, we use the U-Net algorithm to identify the time-frequency GW signals of LIGO O1, O2, and O3 observations. We trained the network using a dataset of pure background noise signals and a combined dataset of GW signals and background noise. In the test dataset, the false alarm probabilities for O1, O2, and O3 were all better than 0.1\%. The trained network was then used to
identify signals in real observations. The time-frequency representations of the GW signals in O1 and O2 were all identified. For O3, the trained network could identify approximately 80\% of GW events. Our results show that the U-Net algorithm can rapidly identify the time-frequency representations of GW signals from BBH mergers.

\acknowledgments
This research has made use of data or softwares obtained from the Gravitational Wave Open Science Center (gwosc.org), a service of LIGO Laboratory, the LIGO Scientific Collaboration, the Virgo Collaboration, and KAGRA.
We thank He Wang for helpful discussions. This work was supported by the National SKA Program of China (2022SKA0110200, 2022SKA0110203), the National Natural Science Foundation of China (12473001, 11975072, 11875102, 11835009), and the National 111 Project (B16009).

\bibliography{unet_gw}

\end{document}